\documentclass[sigconf]{acmart}

\usepackage{wrapfig}
\usepackage[table]{xcolor}
\usepackage{amsmath,amssymb,amsfonts}
\usepackage{algorithmic, algorithm}
\usepackage{graphicx}
\usepackage{textcomp}
\usepackage{xspace}
\usepackage{enumitem}
\usepackage{adjustbox,multirow}
\usepackage{caption}
\usepackage[labelfont=bf,skip=10pt, figurename=Fig.]{caption}
\usepackage{subcaption}
\usepackage{listings}
\usepackage[dvipsnames]{xcolor}
\usepackage{xcolor}
\usepackage[T1]{fontenc}
\usepackage{tikz}
\usepackage{booktabs}
\usepackage{float}
\usepackage{graphicx}
\usepackage{subfiles}
\usepackage{hyperref}
\usepackage[capitalize,nameinlink]{cleveref}
\usepackage{soul}
\usepackage{lipsum}
\usepackage{tabularx}
\usepackage{mwe}
\usepackage{makecell}
\usepackage{tcolorbox}
\usepackage{ulem}
\usepackage{array,booktabs,xcolor}
\newcolumntype{L}[1]{>{\raggedright\arraybackslash}p{#1}}
\newcolumntype{M}[1]{>{\centering\arraybackslash}m{#1}}

\usepackage{pifont}
\newcommand{\cmark}{\textcolor[rgb]{0,0.6,0}{\scalebox{0.85}{\Large\ding{51}}}}
\newcommand{\xmark}{\textcolor[rgb]{0.8,0,0}{\scalebox{0.85}{\Large\ding{55}}}}

\newcommand{\eat}[1]{}

\definecolor{light-gray}{gray}{0.95} 
\definecolor{comment-gray}{rgb}{0.41,0.6,0.33} 
\definecolor{strongred}{RGB}{190, 0, 0}

\lstdefinestyle{mystyle}{
    language=Python, 
    captionpos=b,  
    basicstyle=\fontfamily{phv}\selectfont\footnotesize,  
    keywordstyle=\bfseries,  
    breaklines=true, 
    breakatwhitespace=true, 
    showstringspaces=false, 
    numbers=left, 
    numberstyle=\tiny\color{black}, 
    numbersep=10pt, 
    commentstyle=\color{comment-gray},  
    backgroundcolor=\color{light-gray},  
    xleftmargin=17pt, 
    xrightmargin=3.4pt, 
    framextopmargin=15pt,  
    framexbottommargin=10pt,  
    framexleftmargin=17pt,  
    framesep=3pt, 
    fillcolor=\color{light-gray},  
    linewidth=\linewidth, 
    basewidth=0.5em,
    deletekeywords={[2]file},
    morekeywords={match, if, then, label_host, label_file, drop, alert, allow, declassify, endorse, contains}  
}

\lstdefinestyle{plaintext}{
    basicstyle=\ttfamily\footnotesize,  
    breaklines=true,                    
    breakatwhitespace=true,             
    showstringspaces=false,             
    frame=single,                       
    backgroundcolor=\color{light-gray}, 
    xleftmargin=17pt, 
    xrightmargin=3.4pt, 
    framextopmargin=5pt, 
    framexbottommargin=5pt, 
    framexleftmargin=17pt, 
    framesep=3pt, 
    fillcolor=\color{light-gray}, 
    linewidth=\linewidth,
    basewidth=0.5em,
}

\crefname{figure}{Fig.}{Figs.}
\crefname{listing}{Policy}{Policies}
\crefname{section}{Section}{Sections}
\crefname{table}{Table}{Tables}
\crefname{BNF}{Grammar}{Grammars}
\crefname{algorithm}{Algorithm}{Algorithms}

\hypersetup{%
	bookmarksnumbered, bookmarksopen=true, bookmarksopenlevel=1,%
      colorlinks = false,
}

\AtBeginEnvironment{appendices}{\crefalias{section}{appendix}\crefalias{subsection}{appendix}}

\newif\ifshowcomment
\showcommenttrue

\ifshowcomment
    
    \newcommand{\cici}[1] {{\footnotesize\color{Aquamarine}[CiCi: #1]}}
    \newcommand{\pgao}[1] {{\footnotesize\color{red}[Peng: #1]}}
    \newcommand{\dawn}[1] {{\footnotesize\color{blue}[Dawn: #1]}}
    
\else
    \newcommand{\cici}[1]{}
    \newcommand{\pgao}[1]{}
    \newcommand{\dawn}[1]{}
    
\fi

\newcommand{\cticonnect}{\textsc{CTIConnect}\xspace}

\copyrightyear{2026}
\acmYear{2026}
\setcopyright{cc}
\setcctype{by}
\acmConference[KDD '26]{Proceedings of the 32nd ACM SIGKDD Conference on Knowledge Discovery and Data Mining V.2}{August 09--13, 2026}{Jeju Island, Republic of Korea}
\acmBooktitle{Proceedings of the 32nd ACM SIGKDD Conference on Knowledge Discovery and Data Mining V.2 (KDD '26), August 09--13, 2026, Jeju Island, Republic of Korea}
\acmDOI{10.1145/3770855.3817527}
\acmISBN{979-8-4007-2259-2/2026/08}
\acmSubmissionID{283}

\begin{document}

\title{CTIConnect: A Benchmark for Retrieval-Augmented LLMs over Heterogeneous Cyber Threat Intelligence}

\author{Yutong Cheng}
\affiliation{%
  \institution{Virginia Tech}
  \department{Department of Computer Science}
  \city{Blacksburg}
  \state{VA}
  \country{USA}}
\email{yutongcheng@vt.edu}

\author{Yang Liu}
\affiliation{%
  \institution{Virginia Tech}
  \department{Department of Computer Science}
  \city{Blacksburg}
  \state{VA}
  \country{USA}}
\email{yangliu07@vt.edu}

\author{Changze Li}
\affiliation{%
  \institution{Virginia Tech}
  \department{Department of Computer Science}
  \city{Blacksburg}
  \state{VA}
  \country{USA}}
\email{changzeli@vt.edu}

\author{Dawn Song}
\affiliation{%
  \institution{University of California, Berkeley}
  \department{Department of Computer Science}
  \city{Berkeley}
  \state{CA}
  \country{USA}}
\email{dawnsong@cs.berkeley.edu}

\author{Peng Gao}
\affiliation{%
  \institution{Virginia Tech}
  \department{Department of Computer Science}
  \city{Blacksburg}
  \state{VA}
  \country{USA}}
\email{penggao@vt.edu}

\renewcommand{\shortauthors}{Cheng et al.}

\begin{abstract}
Cyber Threat Intelligence (CTI) is foundational to modern cybersecurity, enabling organizations to proactively defend against evolving threats. However, the sheer volume and heterogeneity of CTI data, spanning structured knowledge bases (CVE, CWE, CAPEC, MITRE ATT\&CK) and unstructured threat reports, far exceed the capacity of manual analysis. The strong contextual understanding and reasoning capabilities of Large Language Models (LLMs) have driven growing research interest in applying them to CTI tasks. 
Yet no existing benchmark evaluates LLMs in a retrieval-augmented setting with a proper evaluation harness—one that grants access to the heterogeneous domain knowledge sources analysts rely on in practice.
To address this gap, we present \textsc{CTIConnect}, a benchmark for systematically evaluating retrieval-augmented LLMs across the CTI task landscape. 
We construct a unified evaluation environment integrating five heterogeneous CTI sources into 1,860 expert-verified QA pairs spanning nine tasks across three categories: Entity Linking, Multi-Document Synthesis, and Entity Attribution. 
Extensive experiments on ten state-of-the-art LLMs reveal that the cross-source semantic gap manifests differently across task categories, demanding fundamentally different retrieval strategies, and that the performance bottleneck shifts between retrieval infrastructure and evidence utilization depending on the task. Our domain-specific strategies further outperform stronger general-purpose retrieval paradigms (retrieve-then-rerank, IRCoT), showing that closing this gap requires structural interventions rather than generic retrieval improvements. These findings hold across all ten LLMs, remain consistent on the full benchmark, and stay stable under temporal splits spanning 2008--2025. Together, they provide actionable guidance for designing scalable knowledge retrieval architectures over large-scale, heterogeneous CTI ecosystems.
\end{abstract}

\ccsdesc[300]{Information systems~Specialized information retrieval}
\ccsdesc[300]{Security and privacy~Intrusion/anomaly detection and malware mitigation}

\keywords{Cyber Threat Intelligence, Large Language Models, Benchmark, Security}

\maketitle


\section{Introduction}

\begin{figure*}[!t]
    \centering
     \includegraphics[width=0.99\linewidth]{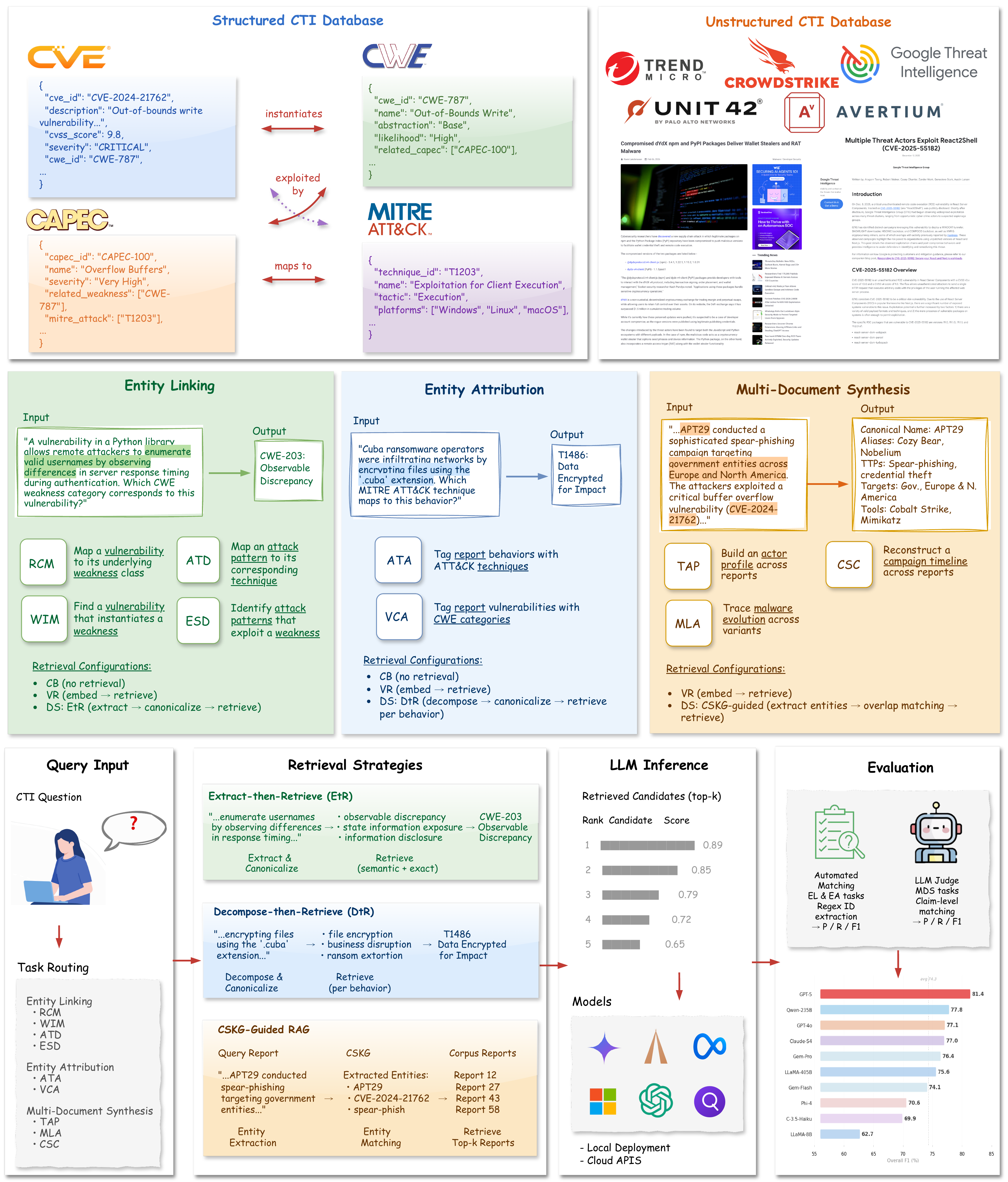}
             \vspace{-2ex}
\caption{Overview of \textsc{CTIConnect}, which integrates five heterogeneous CTI sources (CVE, CWE, CAPEC, MITRE ATT\&CK, vendor threat reports), nine tasks across three categories (Entity Linking, Entity Attribution, Multi-Document Synthesis), and a unified evaluation pipeline comparing domain-specific retrieval strategies against vanilla RAG baselines.}
    \label{fig:framework}
        \vspace{-3ex}
\end{figure*}

Cyber Threat Intelligence (CTI) is foundational to modern cybersecurity defense, enabling organizations to shift from reactive incident response to proactive threat anticipation~\cite{10117505}. However, the CTI data ecosystem is both massive and heterogeneous, spanning structured knowledge bases (CVE, CWE, CAPEC, MITRE ATT\&CK) and unstructured vendor threat reports that employ fundamentally different schemas, vocabularies, and abstraction levels. The volume and heterogeneity of this ecosystem far exceed the capacity of manual analysis to scale.
The contextual understanding and reasoning capabilities of Large Language Models (LLMs) position them as a natural candidate for addressing this challenge, driving a growing body of research on LLM-powered CTI analysis, from extracting structured threat information from narrative reports~\cite{intelex, ttprint, gao2021enabling, towardsnguyen} to constructing comprehensive cybersecurity knowledge graphs~\cite{ctinexus, gao2024threatkg1, tikg}. Industry interest is also growing: platforms such as Microsoft Security Copilot~\cite{microsoft_copilot} now integrate LLMs into operational CTI workflows for real-time threat analysis.

\begin{table*}[t]
\centering
\small
    \vspace{-0.2cm}
\caption{Task coverage comparison. CTIConnect covers all nine CTI tasks across three task categories, while existing benchmarks address only subsets.}
    \vspace{-0.2cm}
\label{tab:task-coverage}
\begin{tabular}{ll|ccc}
\toprule
\textbf{Task Category} & \textbf{Task} & \textbf{CTIBench~\cite{ctibench}} & \textbf{SEvenLLM~\cite{sevenllm}} & \textbf{CTIConnect} \\
\midrule
\multirow{4}{*}{\textit{Entity Linking}}
& RCM: Root Cause Mapping & \cmark & \xmark & \cmark \\
& WIM: Weakness Instantiation & \xmark & \xmark & \cmark \\
& ATD: Attack Technique Derivation & \cmark & \xmark & \cmark \\
& ESD: Exploitation Surface Discovery & \xmark & \xmark & \cmark \\
\midrule
\multirow{3}{*}{\textit{Multi-Doc Synthesis}}
& MLA: Malware Lineage Analysis & \xmark & \cmark & \cmark \\
& TAP: Threat Actor Profiling & \cmark & \cmark & \cmark \\
& CSC: Campaign Storyline Construction & \xmark & \xmark & \cmark \\
\midrule
\multirow{2}{*}{\textit{Entity Attribution}}
& VCA: Vulnerability Catalog Attribution & \xmark & \xmark & \cmark \\
& ATA: Attack Technique Attribution & \xmark & \xmark & \cmark \\
\midrule
\multicolumn{2}{c|}{\textbf{Total Coverage}} & \textbf{3/9} & \textbf{2/9} & \textbf{9/9} \\
\bottomrule
\end{tabular}
    \vspace{-0.3cm}
\end{table*}

As interest in applying LLMs to CTI analysis grows, rigorous benchmarks become essential for evaluating model performance in this domain. Yet existing CTI benchmarks address only a fraction of this need: CTIBench~\cite{ctibench} tests parametric knowledge under closed-book settings, and SEvenLLM~\cite{sevenllm} measures gains from domain-specific fine-tuning. Neither paradigm reflects how LLMs are deployed for CTI in practice. CTI knowledge evolves rapidly: new vulnerabilities are disclosed daily, threat reports are published continuously, and adversary behaviors shift faster than any retraining cycle can follow. The volume of accumulated domain knowledge far exceeds what parametric memory can internalize, yet factual precision remains critical for downstream detection and response. These characteristics make retrieval-augmented generation (where models actively retrieve and reason over external knowledge at inference time) a prerequisite for any production CTI system. Yet \textbf{no} existing benchmark evaluates this capability, leaving a fundamental question unanswered: \textit{What retrieval strategies enable LLMs to most effectively leverage external knowledge across the heterogeneous, multi-source CTI ecosystem?}

\vspace{+0.7cm}
Answering this question is non-trivial. General-purpose RAG benchmarks~\cite{rgb, crag, ragbench, multihoprag} typically evaluate retrieval over homogeneous corpora, but the CTI data ecosystem is fundamentally heterogeneous: it spans structured knowledge bases that encode information in formal, technique-oriented terminology (e.g., \texttt{T1003.001 -- LSASS Memory}) and unstructured vendor reports that describe the same behaviors in analyst-authored narrative prose (e.g., ``the adversary harvested credentials from memory''). This heterogeneity creates a \textit{cross-source semantic gap}: queries expressed in one vocabulary systematically fail to match relevant evidence encoded in another, causing vanilla embedding-based retrieval to break down. The gap is further compounded by the scale and complexity of the CTI ecosystem: four major knowledge bases~\cite{CVE, CWE, CAPEC, mitre} maintain hundreds of thousands of cross-source mappings, while reports from dozens of vendors reference the same entities under inconsistent aliases (e.g., the same Russian state-sponsored threat group appears as ``APT29,'' ``Cozy Bear,'' or ``Nobelium'' depending on the vendor). A concrete illustration of why such correlation is operationally valuable is the \emph{Magniber} ransomware family: cross-source analysis links a 2023 SmartScreen-bypass campaign to a 2021 PrintNightmare-based variant and a 2017 South Korea targeting campaign, revealing a six-year operational continuity invisible to single-report analysis (\cref{app:magniber_case_study}). No existing benchmark captures these cross-source retrieval challenges, making it impossible to assess whether current retrieval strategies are adequate for production CTI systems.

To address these challenges, we present \textsc{CTIConnect}, a benchmark for systematically evaluating retrieval strategies for LLM-based CTI analysis. We make two complementary contributions. First, we construct a \textit{unified heterogeneous CTI evaluation environment} integrating five major knowledge bases (CVE, CWE, CAPEC, MITRE ATT\&CK, and vendor threat reports) into nine tasks across three categories (Entity Linking, Multi-Document Synthesis, Entity Attribution) that cover all cross-source directions in CTI analysis (Table~\ref{tab:task-coverage}). 
Second, we design \textit{domain-specific retrieval strategies} tailored to each task category's semantic gap characteristics and evaluate them against vanilla RAG baselines across ten LLMs spanning open-source and proprietary families. 
Experiments show that domain-specific strategies yield substantial improvements over vanilla RAG (up to +35.2\% for entity linking, +16.0\% for attribution, and +11.3\% for synthesis), with optimal strategies varying by task category. 
We further show that this advantage cannot be replicated by stronger general-purpose retrieval paradigms: retrieve-then-rerank and IRCoT recover at most 1--5\% of the gap that our domain-specific strategies close, confirming that bridging the cross-source gap requires structural interventions rather than generic retrieval improvements.
These findings hold across ten LLMs, remain consistent on the full 1,860-pair benchmark, and stay stable under per-task temporal splits across the 2008--2025 span.
In summary, this paper makes the following contributions:

\begin{figure*}[!t]
    \centering
        \vspace{-0.2cm}
    \includegraphics[width=\linewidth]{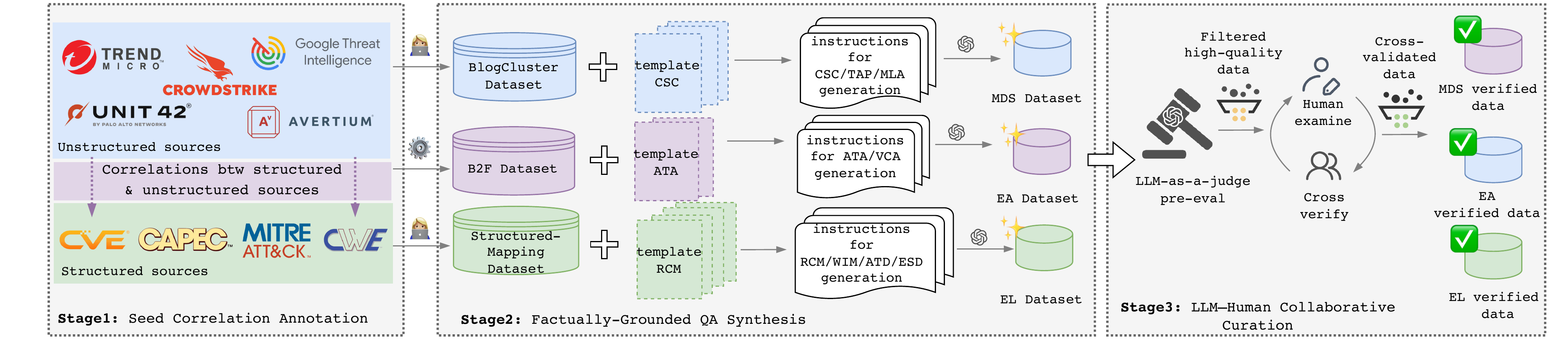}
        \vspace{-0.4cm}
    \caption{Benchmark construction pipeline. Each CTI task is created through three stages: cross-source seed annotation from authoritative CTI databases, template-constrained QA synthesis, and LLM-human collaborative curation for quality control.}
    \label{fig:data_construct}
    \vspace{-0.3cm}
\end{figure*}

\begin{itemize}[leftmargin=*]
\item \textbf{Benchmark.} We construct and release \textsc{CTIConnect}, the first retrieval-augmented evaluation environment for CTI analysis, integrating structured knowledge bases (CVE, CWE, CAPEC, MITRE ATT\&CK) and unstructured threat reports from 35 sources into 1,860 expert-verified QA pairs spanning nine core security analysis tasks across three categories.

\item \textbf{Retrieval Strategies.} We design domain-specific retrieval strategies that bridge the distinct semantic gap characteristics of each task category, and contrast them with vanilla RAG as well as stronger general-purpose paradigms.

    \item \textbf{Findings.} Systematic evaluation of ten LLMs reveals that the cross-source semantic gap undermines both retrieval accuracy and evidence utilization to varying degrees across task categories. Our analysis examines the failure modes of current retrieval strategies and provides insights for future RAG improvements in CTI systems. We release \textsc{CTIConnect} at \href{https://cticonnect.github.io/}{\textcolor{purple}{cticonnect.github.io}} and invite the community to evaluate new approaches on our benchmark and to build upon our retrieval-strategy findings.
\end{itemize}

\section{Related Work}

\paragraph{Cyber Threat Intelligence}
Cyber Threat Intelligence (CTI) is evidence-based knowledge about existing or emerging cyber threats that enables organizations to anticipate and defend against adversary activities~\cite{gartner_cti, liao2016acing}. Such intelligence ultimately supports
downstream defensive operations such as attack investigation, which
analyzes large-scale system provenance to reconstruct multi-step
attacks like APTs~\cite{hitl}.
\textbf{Structured CTI} comprises interconnected knowledge bases: CVE~\cite{CVE} catalogs vulnerabilities, CWE~\cite{CWE} captures underlying weakness patterns, CAPEC~\cite{CAPEC} documents attack patterns, and MITRE ATT\&CK~\cite{mitre} organizes adversary techniques. These sources encode knowledge at varying abstraction levels using formal, technique-oriented terminology.
\textbf{Unstructured CTI} consists of vendor threat reports (e.g., CrowdStrike~\cite{CrowdStrike}, Unit 42~\cite{Unit42}, Trend Micro~\cite{trendmicro}) that describe adversary activities in narrative form. These sources are inherently noisy: actionable intelligence such as threat actor identities, malware lineage, and campaign progression is buried within lengthy prose, and the same entity may surface under different aliases across vendors (e.g., ``APT29,'' ``Cozy Bear,'' ``Nobelium''). Together, these heterogeneous sources form a fast-evolving, knowledge-intensive ecosystem whose scale and update velocity make retrieval-augmented approaches essential for production CTI systems.

\paragraph{CTI Benchmarks}
Although LLM benchmarks have proliferated in general cybersecurity~\cite{secbench, lee2025secbenchautomatedbenchmarkingllm, shao2025nyuctfbenchscalable, zhang2024cybench, zhu2025cve, ullah2024llms, CyberMetric}, only a few have targeted the CTI domain.
CTIBench~\citep{ctibench} was a notable effort, evaluating LLMs on four tasks: root cause mapping, vulnerability severity prediction, attack technique extraction, and threat actor attribution.
However, these tasks capture only a narrow slice of the CTI analysis landscape.
CTIBench also evaluated only a few models under closed-book settings, relying solely on parametric knowledge without retrieval or augmentation.
Its dataset was manually curated at small scale, making it difficult to extend to emerging threats.
Another effort, SEvenLLM~\citep{sevenllm}, introduced SEvenLLM-Bench, a bilingual multi-task dataset covering 28 CTI-related tasks (13 understanding and 15 generation).
The benchmark is restricted to single-report extraction and summarization, and limits evaluation to instruction-tuned small models ($\leq$14B parameters), excluding frontier LLMs that may exhibit stronger reasoning capabilities.

\paragraph{LLMs for Security}
The rapid advancement of LLMs has catalyzed growing research across diverse cybersecurity domains, spanning vulnerability detection~\cite{scale_vuldetect, vulllm}, fuzz testing~\cite{gptfuzzer, whitefox}, reverse engineering~\cite{latte_binary}, intrusion detection~\cite{liu2023malurl}, penetration testing~\cite{pentestgpt}, and smart contract analysis~\cite{gptlens}.
Among these domains, \emph{CTI analysis} is particularly knowledge-intensive. Early work extracted structured information from narrative reports, such as TTPs~\cite{ttprint, Bchel2025SoKAT} and STIX bundles~\cite{action}; later studies moved to relation modeling and cybersecurity knowledge graph construction~\cite{ctinexus, tikg, alam2023lookingiocsautomaticallyextracting}. More recent efforts have applied LLMs to downstream CTI operations such as automated detection rule extraction from cloud-based threat intelligence~\cite{llmcloudhunter} and agentic CTI workflows that orchestrate multi-step analysis pipelines~\cite{cylens}. However, these efforts share a common paradigm: extracting useful knowledge from CTI sources or training models to internalize domain knowledge. We approach the problem from a complementary angle: evaluating whether LLMs can serve as an intermediate reasoning layer that connects analysts with the full heterogeneous CTI ecosystem, by retrieving and integrating cross-source knowledge on demand.

\section{\cticonnect{} Design}
\label{sec:design}

\cticonnect{} systematically evaluates retrieval strategies for LLM-based CTI analysis. We first define the task landscape that the benchmark covers (\S\ref{sec:task-design}), then describe the heterogeneous data sources that support each task category and the construction process that ensures data quality (\S\ref{sec:benchmark-data}).

\subsection{Task Design}
\label{sec:task-design}

CTI analysis operates over heterogeneous data spanning structured knowledge bases and unstructured threat reports. We organize the core cross-source operations into three task categories based on the source types they bridge: \textit{Entity Linking} (structured $\rightarrow$ structured) aligns entries across CTI taxonomies; \textit{Entity Attribution} (unstructured $\rightarrow$ structured) grounds narrative descriptions to formal taxonomy entries; and \textit{Multi-Document Synthesis} (unstructured $\rightarrow$ unstructured) aggregates intelligence scattered across vendor reports. Together, these categories cover all cross-source directions over heterogeneous CTI data, enabling systematic evaluation of retrieval strategies under distinct semantic gap conditions.

\subsubsection{Entity Linking}

Entity linking tasks map an entry in one structured CTI knowledge base to its corresponding entry in another. The retrieval space is vast: the model must discriminate among 900+ CWE categories, 200K+ CVE records, or 500+ CAPEC patterns to identify a correct match. In addition, subtle distinctions between semantically similar entries (e.g., CWE-787 \textit{Out-of-bounds Write} vs.\ CWE-125 \textit{Out-of-bounds Read}) demand precise vocabulary alignment, making canonicalization the primary retrieval bottleneck.

We define four tasks spanning the major CTI knowledge bases:
\ding{182}~\textbf{CTI-RCM} (Root Cause Mapping, CVE $\rightarrow$ CWE) identifies the underlying weakness category of a disclosed vulnerability from its CVE description, mapping terse, vendor-specific language to the correct abstract weakness class.
\ding{183}~\textbf{CTI-WIM} (Weakness Instantiation Mapping, CWE $\rightarrow$ CVE) reverses this direction: given an abstract weakness description, the model identifies a concrete CVE that instantiates it.
\ding{184}~\textbf{CTI-ATD} (Attack Technique Derivation, CAPEC $\rightarrow$ ATT\&CK) maps a CAPEC attack pattern to its corresponding MITRE ATT\&CK technique, aligning two overlapping but independently maintained taxonomies with differing granularity.
\ding{185}~\textbf{CTI-ESD} (Exploitation Surface Discovery, CWE $\rightarrow$ CAPEC) reasons from the defensive perspective (weakness) to the offensive perspective (attack pattern), identifying how a given weakness can be exploited.

\subsubsection{Entity Attribution}

Entity attribution tasks ground narrative attack or vulnerability descriptions in threat reports to formal entries in structured CTI taxonomies. Analyst-authored narratives employ action-oriented language (e.g., ``the adversary harvested credentials from memory'') that differs fundamentally from technique-oriented taxonomy terminology (e.g., \texttt{T1003.001\,--\,LSASS Memory}), creating a vocabulary mismatch that generic embeddings cannot bridge. The one-to-many nature further compounds the difficulty: a single passage may describe multiple interleaved behaviors, requiring decomposition into atomic actions, independent retrieval per behavior, and aggregation of results.

\ding{182}~\textbf{CTI-ATA} (Attack Technique Attribution, Report $\rightarrow$ ATT\&CK entries) identifies all MITRE ATT\&CK techniques described in a threat report passage by decomposing complex behavioral narratives into atomic actions, even when a single sentence conflates multiple tactics (e.g., lateral movement \textit{and} credential access).
\ding{183}~\textbf{CTI-VCA} (Vulnerability Catalog Attribution, Report $\rightarrow$ CWE entries) identifies all CWE weakness categories referenced in a vulnerability exploitation narrative. The challenge is amplified by indirection: reports describe exploitation \textit{effects} rather than naming weaknesses directly, requiring the model to infer root causes from observable consequences.

\subsubsection{Multi-Document Synthesis}

Multi-document synthesis tasks aggregate intelligence about the same entity scattered across multiple vendor threat reports. Unlike entity attribution where vocabulary mismatch is the primary barrier, synthesis faces a distinct challenge: entity aliasing. Different vendors describe the same threat actor, malware family, or campaign under inconsistent naming conventions (e.g., ``APT29,'' ``Cozy Bear,'' and ``Nobelium''), creating near-miss distractors that outscore gold documents in embedding space and producing the largest semantic gap of any task category.

We define three tasks along distinct analytical dimensions:
\ding{182}~\textbf{CTI-TAP} (Threat Actor Profiling) synthesizes a comprehensive actor profile (covering TTPs, targets, and toolsets) from reports referencing the same group under different aliases, requiring implicit entity resolution before aggregation.
\ding{183}~\textbf{CTI-MLA} (Malware Lineage Analysis) traces the evolutionary lineage of a malware family across reports on related variants, identifying capability progression and code reuse through temporal reasoning over independently authored and potentially contradictory accounts.
\ding{184}~\textbf{CTI-CSC} (Campaign Storyline Construction) reconstructs a coherent campaign timeline from reports documenting different phases of the same operation, reconciling fragmented, partially overlapping accounts into a unified narrative with consistent chronology.

\begin{table}[t]
\centering
\small
\caption{Benchmark data summary. \textsc{CTIConnect} comprises 1,860 expert-verified QA pairs spanning nine tasks across three data sources, distinguished by their ground-truth characteristics and produced through a single unified construction pipeline (\S\ref{sec:benchmark-data}).}
    \vspace{-0.2cm}
\label{tab:data-summary}
\begin{tabular}{l|l|r|l}
\toprule
\textbf{Data Source} & \textbf{Task} & \textbf{QA} & \textbf{Ground Truth} \\
\midrule
\multirow{4}{*}{Structured Mappings}
 & RCM & 290 & \multirow{4}{*}{Official KB links} \\
 & WIM & 308 & \\
 & ATD & 261 & \\
 & ESD & 280 & \\
\midrule
\multirow{3}{*}{Report Clusters}
 & TAP & 135 & \multirow{3}{*}{Manual clustering} \\
 & MLA & 95  & \\
 & CSC & 111 & \\
\midrule
\multirow{2}{*}{B2F Alignments}
 & ATA & 160 & \multirow{2}{*}{Expert annotation} \\
 & VCA & 220 & \\
\midrule
\textbf{Total} & \textbf{9 tasks} & \textbf{1,860} & --- \\
\bottomrule
\end{tabular}
\vspace{-0.4cm}
\end{table}

\begin{table*}[!t]
\centering
    \vspace{-0.2cm}
\caption{Performance of ten LLMs across nine CTI tasks under three retrieval configurations, evaluated on a 691-pair subset of \cticonnect{}.$^\dagger$ Results on the full 1,860-pair benchmark (three representative models) are reported in~\cref{sec:scalability}.}
    \vspace{-0.2cm}
\label{tab:main_result}
\resizebox{\textwidth}{!}{
\begin{tabular}{l|ccc|ccc|ccc|ccc|cc|cc|cc|ccc|ccc}
\toprule
\multirow{3}{*}{\textbf{Model}}
& \multicolumn{12}{c|}{\textbf{Entity Linking }}
& \multicolumn{6}{c|}{\textbf{Multi-Doc Synthesis }}
& \multicolumn{6}{c}{\textbf{Entity Attribution }} \\
\cmidrule(lr){2-13} \cmidrule(lr){14-19} \cmidrule(lr){20-25}
& \multicolumn{3}{c|}{\textit{RCM}} & \multicolumn{3}{c|}{\textit{WIM}} & \multicolumn{3}{c|}{\textit{ATD}} & \multicolumn{3}{c|}{\textit{ESD}}
& \multicolumn{2}{c|}{\textit{CSC}} & \multicolumn{2}{c|}{\textit{TAP}} & \multicolumn{2}{c|}{\textit{MLA}}
& \multicolumn{3}{c|}{\textit{ATA}} & \multicolumn{3}{c}{\textit{VCA}} \\
\cmidrule(lr){2-4} \cmidrule(lr){5-7} \cmidrule(lr){8-10} \cmidrule(lr){11-13}
\cmidrule(lr){14-15} \cmidrule(lr){16-17} \cmidrule(lr){18-19}
\cmidrule(lr){20-22} \cmidrule(lr){23-25}
& CB & VR & DS & CB & VR & DS & CB & VR & DS & CB & VR & DS
& VR & DS & VR & DS & VR & DS
& CB & VR & DS & CB & VR & DS \\
\midrule
\rowcolor{blue!8} \multicolumn{25}{c}{\textit{Open-Source Models}} \\
\midrule
LLaMA-3-405B
& .14 & .68 & .98 & .01 & .62 & .95 & .06 & .71 & .99 & .01 & .65 & .97
& .56 & .67 & .43 & .65 & .31 & .32
& .58 & .56 & .63 & .42 & .54 & .64 \\
LLaMA-3-8B
& .05 & .56 & .91 & .00 & .50 & .84 & .02 & .59 & .93 & .00 & .54 & .89
& .48 & .59 & .46 & .42 & .26 & .38
& .13 & .28 & .28 & .10 & .22 & .40 \\
Phi-4
& .07 & .63 & .95 & .00 & .57 & .90 & .03 & .66 & .96 & .00 & .61 & .95
& .51 & .62 & .55 & .76 & .40 & .36
& .31 & .38 & .49 & .24 & .48 & .36 \\
Qwen-3-235B
& .10 & \underline{.73} & \textbf{1.0} & .01 & \underline{.68} & .98 & .07 & \underline{.75} & \textbf{1.0} & .00 & \underline{.70} & \textbf{1.0}
& .66 & \underline{.71} & .58 & .71 & .30 & .41
& .58 & .58 & \underline{.65} & .40 & .56 & .54 \\
\midrule
\rowcolor{red!8} \multicolumn{25}{c}{\textit{Proprietary Models}} \\
\midrule
GPT-4o
& .08 & .69 & .98 & .01 & .63 & .96 & .04 & .72 & \underline{.99} & .00 & .67 & .98
& .66 & .66 & .58 & .67 & .36 & .39
& .60 & .64 & .69 & .48 & .58 & .62 \\
GPT-5
& .15 & \textbf{.75} & .99 & .03 & \textbf{.70} & .97 & .09 & \textbf{.76} & .99 & \textbf{.02} & \textbf{.72} & \textbf{1.0}
& \textbf{.72} & .67 & .58 & .66 & .36 & .39
& \textbf{.83} & .74 & \textbf{.90} & \textbf{.64} & .60 & \textbf{.76} \\
Gemini-2.5-Pro
& .06 & .65 & .96 & .02 & .59 & .93 & \textbf{.11} & .68 & .98 & .00 & .63 & .98
& .61 & .61 & .53 & \textbf{.79} & .33 & .36
& \underline{.63} & .56 & .57 & .58 & \textbf{.74} & .70 \\
Gemini-2.5-Flash
& .04 & .64 & .96 & .01 & .58 & .92 & .06 & .67 & \underline{.99} & .00 & .62 & .97
& .57 & .69 & .54 & .61 & .32 & \textbf{.44}
& .53 & .47 & .55 & .56 & .60 & .54 \\
Claude-Sonnet-4
& \textbf{.26} & .72 & \underline{.99} & \textbf{.08} & .67 & \textbf{1.0} & .05 & .73 & \underline{.99} & \underline{.01} & .69 & \underline{.99}
& .48 & .55 & .51 & .56 & .41 & \underline{.48}
& .71 & .67 & .73 & .46 & .58 & \underline{.64} \\
Claude-3.5-Haiku
& .09 & .61 & .95 & .02 & .55 & .92 & .03 & .64 & .97 & .00 & .59 & .96
& .44 & .55 & .48 & .58 & \underline{.44} & .41
& .47 & .48 & .49 & .50 & .52 & .46 \\
\midrule
\multicolumn{1}{c|}{\textbf{Average}}
& .10 & .67 & .97 & .02 & .61 & .94 & .06 & .69 & .98 & .00 & .64 & .97
& .57 & .63 & .52 & .64 & .35 & .39
& .54 & .54 & .60 & .44 & .54 & .57 \\
\bottomrule
\end{tabular}
    \vspace{-0.3cm}
}
\raggedright\footnotesize{
$^\dagger$ \textbf{Bold}: best; \underline{underline}: second-best per column. CB = Closed-Book; VR = Vanilla RAG; DS = Domain-Specific strategy (EtR for Entity Linking, CSKG-guided for Multi-Doc Synthesis, DtR for Entity Attribution). Multi-Doc Synthesis omits CB as the task inherently requires multi-document retrieval.
}
    \vspace{-0.3cm}
\end{table*}

\subsubsection{Structured Cross-Source Mappings}
\textit{(Supporting Entity Linking tasks.)} The four entity linking tasks rely on authoritative mappings maintained by security standards across interconnected CTI knowledge bases: CVE$\rightarrow$CWE (vulnerability to root cause), CWE$\rightarrow$CVE (weakness to instantiation), CWE$\rightarrow$CAPEC (weakness to attack pattern), and CAPEC$\rightarrow$ATT\&CK (attack pattern to adversary technique). These mappings are curated by MITRE and NVD, encompassing 200K+ CVE--CWE pairs and 500+ CAPEC--ATT\&CK relationships. From these mappings we construct 1,139 QA pairs across the four linking tasks, spanning diverse vulnerability types and abstraction levels. As the ground truth derives from the official knowledge bases, factual reliability is inherently guaranteed.

\subsubsection{Blog-to-Framework Alignments}
\textit{(Supporting Entity Attribution tasks.)} The two attribution tasks require ground-truth mappings from narrative passages in threat reports to formal taxonomy entries. We construct a Blog-to-Framework (B2F) dataset that links attack-behavior phrases in threat reports to ATT\&CK techniques or CWE categories, combining direct practitioner annotation with LLM-assisted candidate mapping under expert verification. This process yields 380 QA pairs spanning diverse attack behaviors and vulnerability patterns, with each passage mapped to one or more taxonomy entries.

\subsubsection{Report Clusters}
\textit{(Supporting Multi-Document Synthesis tasks.)} The three synthesis tasks require sets of topically related threat reports about the same entity: a threat actor, malware family, or campaign. We aggregate 321 reports from 35 vendor sources (e.g., CrowdStrike, Unit~42, Trend Micro) and cluster them into adversary-centric groups through topic identification, alias resolution (e.g., linking ``APT29,'' ``Cozy Bear,'' and ``Nobelium'' to the same actor), and metadata reconciliation. Only clusters containing reports from at least two distinct vendors are retained, and from these clusters we derive 341 QA pairs across the three synthesis tasks.

\subsubsection{Construction Pipeline}
\label{sec:benchmark-data}
All three data sources share a unified three-stage construction pipeline (\cref{fig:data_construct}): \ding{182}~\textit{Seed annotation} establishes ground-truth correlations from authoritative sources. For entity linking, these derive directly from official cross-source mappings maintained by MITRE and NVD, ensuring inherent factual reliability. For entity attribution and multi-document synthesis, seed correlations are established through dual expert annotation with senior adjudication. \ding{183}~\textit{Template-constrained QA synthesis} transforms each correlation into task-specific question--answer pairs, with prompt templates specifying the task instruction, input entity, expected output, and required format. By grounding generation in verified correlations, this step reduces hallucination while enabling scalable data production. \ding{184}~\textit{Multi-layered quality control} enforces annotation reliability through three mechanisms: an LLM-based judge (GPT-4) filters low-confidence samples via a structured rubric, two domain practitioners independently verify the remaining pairs, and a senior annotator with over three years of CTI analysis experience conducts final adjudication. The resulting benchmark comprises 1,860 high-quality QA pairs across nine tasks, each grounded in authoritative CTI sources.

\section{Experiments}
\label{sec:experiments}

\begin{figure*}[t]
    \centering
    \includegraphics[width=0.85\linewidth]{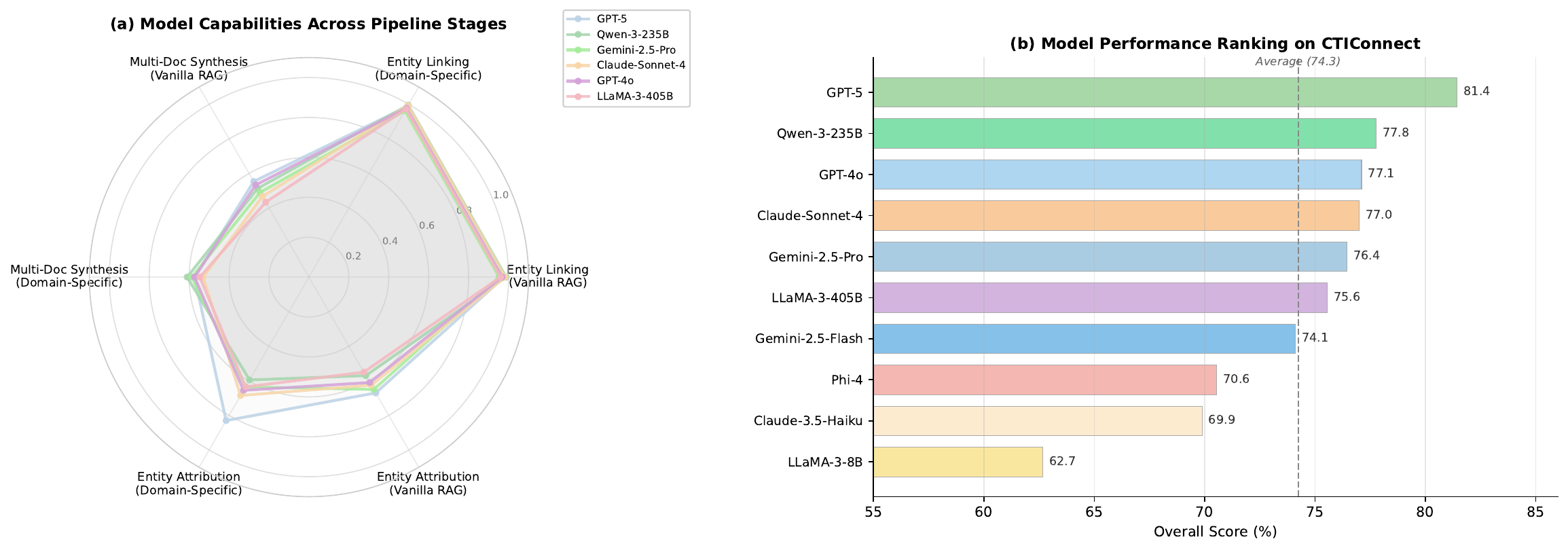}
    \caption{Overall performance on CTIConnect. \textbf{(a)}~Model capabilities across task categories under vanilla RAG and domain-specific retrieval. \textbf{(b)}~Weighted performance ranking of all ten models.}
        \vspace{-0.2cm}
    \label{fig:radar_overall}
\end{figure*}
We conduct extensive experiments to evaluate retrieval strategies across the CTI pipeline. We first describe the experimental setup (\S\ref{sec:setup}), then present overall performance (\S\ref{sec:performance}), followed by detailed analyses including retrieval-paradigm comparisons, full-benchmark verification, and temporal robustness (\S\ref{sec:findings}).

\subsection{Experimental Setup}
\label{sec:setup}

\paragraph{Models}
\looseness=-1We evaluate ten LLMs spanning four proprietary families and three open-source families: GPT-5, GPT-4o (OpenAI); Claude-Sonnet-4, Claude-3.5-Haiku (Anthropic); Gemini-2.5-Pro, Gemini-2.5-Flash (Google); Qwen-3-235B (Alibaba); Phi-4 (Microsoft); LLaMA-3-405B, LLaMA-3-8B (Meta). All models are accessed via their respective APIs with default decoding parameters.

\paragraph{Retrieval Configurations}
All task categories are evaluated under two shared baselines and one domain-specific strategy. \textit{Closed-Book (CB)} provides the performance floor by requiring the LLM to answer using only parametric knowledge, measuring how much CTI knowledge is encoded during pretraining. \textit{Vanilla RAG (VR)} embeds the raw query directly (without any query transformation) and retrieves top-$k$ passages from the knowledge base via cosine similarity against both structured KB entries and unstructured report chunks. These two baselines isolate the contribution of retrieval itself (CB $\rightarrow$ VR) from the contribution of retrieval \textit{strategy} (VR $\rightarrow$ domain-specific). 

Beyond these baselines, we analyze vanilla RAG's failure modes on each task category and find that they stem from distinct manifestations of the cross-source semantic gap (\S\ref{sec:findings-semantic-gap}). Guided by these failure patterns and the structural properties of each data source, we design three domain-specific retrieval baselines:

\begin{itemize}[leftmargin=*]
\item \textit{Extract-then-Retrieve (EtR).} For entity linking, the LLM first extracts the security-relevant semantic content of the input description (e.g., vulnerability type, weakness mechanism, and impact) and canonicalizes it into keyphrases aligned with the target knowledge base. The canonicalized keyphrases are then embedded as a dense query for top-$k$ retrieval against the target KB. By extracting the semantic content from the source-side wording and re-expressing it in the target-side vocabulary before encoding, EtR addresses the source--target vocabulary mismatch underlying the cross-source semantic gap, while keeping the retrieval path itself purely dense.

\item \textit{CSKG-Guided RAG.} For multi-document synthesis, we construct a Cybersecurity Knowledge Graph (CSKG) offline using CTINexus~\cite{ctinexus}: each report is reduced to a sparse bag of canonical entities by extracting STIX-aligned named entities and resolving aliases (e.g., ``APT29''~$\equiv$~``Cozy Bear''~$\equiv$~``Nobelium'') against a MITRE-Groups dictionary. At query time, the input report is processed through the same pipeline, and corpus reports are ranked against the query entity bag via BM25 with IDF weighting; the top-$k$ reports are retrieved as context. By operating on canonical entities rather than raw text chunks, CSKG-Guided RAG bypasses both the entity-aliasing and chunk-noise failure modes of vanilla RAG that we analyze in~\cref{sec:findings-semantic-gap}, and remains robust to the choice of extraction model (\cref{app:cskg_robustness}).

\item \textit{Decompose-then-Retrieve (DtR).} For entity attribution, DtR first decomposes the input passage into $N$ atomic behaviors, canonicalizes each into taxonomy-aligned vocabulary, and performs independent retrieval per behavior before aggregation.
\end{itemize}

\paragraph{Metrics}
For entity linking and entity attribution tasks, we use Precision, Recall, and F1-score after regex-based identifier normalization. For multi-document synthesis tasks, we use GPT-4 as an automatic judge guided by a structured rubric. The judge is comprehensively validated on a 20\% expert-annotated sample, achieving Cohen's $\kappa = 0.85$ with three CTI experts (inter-human $\kappa = 0.93$), 93.9\% self-consistency over five runs, and a stylistic-bias differential of only 0.013 between GPT-family and non-GPT outputs (full protocol and results in~\cref{app:judge_reliability}).

\begin{figure}[t]
    \centering
        \vspace{-0.2cm}
    \includegraphics[width=\linewidth]
    {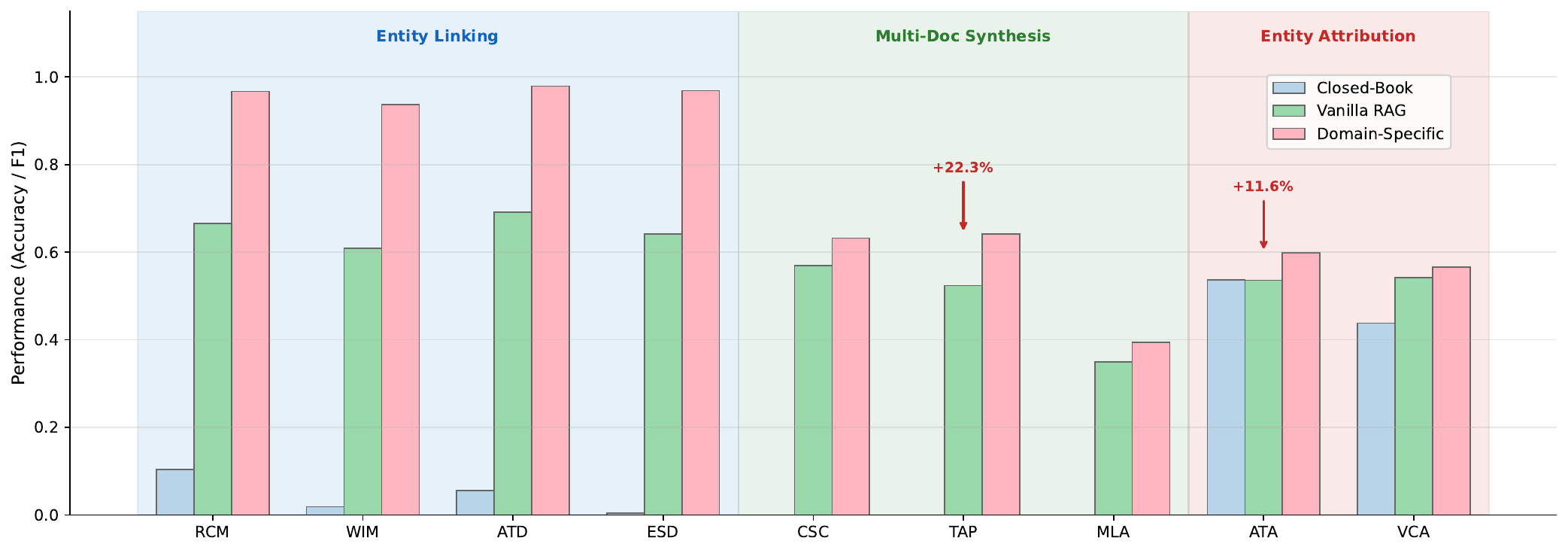}
    \caption{Retrieval strategy comparison across all nine tasks. Average performance under closed-book (red), vanilla RAG (orange), and domain-specific retrieval (green) configurations.}
    \label{fig:retrieval_comparison}
    \vspace{-0.4cm}
\end{figure}

\subsection{Overall Performance on CTIConnect}
\label{sec:performance}

\paragraph{Overview}
Table~\ref{tab:main_result} presents the performance of ten LLMs across all nine tasks under three retrieval configurations, evaluated on a 691-pair subset of \cticonnect{}. With domain-specific retrieval, models average 74.3\% overall, but this aggregate masks stark asymmetries: entity linking saturates near ceiling (average DS: 97\%), while multi-document synthesis (55.3\%) and entity attribution (58.5\%) remain substantially below, indicating that these categories require both better retrieval strategies and stronger model reasoning to close the gap. Meanwhile, the overall model ranking is tight at the top: GPT-5 leads at 81.4\%, with three models clustered within 77--78\%. However, no single model dominates all categories.

\paragraph{Category Performance}
As illustrated in Figure~\ref{fig:radar_overall}~(a), the radar chart reveals a highly asymmetric capability profile across task categories. Entity linking with domain-specific retrieval saturates near ceiling for most models (average DS: 97\%), yet collapses to near-zero under closed-book conditions (average CB: 4.5\%), confirming that these tasks primarily test retrieval rather than parametric knowledge. Multi-document synthesis remains moderately challenging (average:55.3\%), with performance heavily dependent on retrieval coverage across vendor reports. Entity attribution exhibits the widest variance (28\%--90\%), suggesting that this category most effectively discriminates model capabilities.

\paragraph{Retrieval Strategy Impact}
Figure~\ref{fig:retrieval_comparison} compares retrieval strategies across all nine tasks. Domain-specific strategies outperform vanilla RAG in every task category, but the mechanism differs by retrieval structure: lexical canonicalization via LLM query rewriting for entity linking, alias resolution for multi-document synthesis, and decomposition plus canonicalization for entity attribution. For entity linking, EtR delivers a large average VR$\rightarrow$DS gain of $\sim$31 percentage points (e.g., LLaMA-3-8B WIM: .50$\rightarrow$.84; Qwen-3-235B RCM: .73$\rightarrow$1.0): under vanilla RAG, gold entries sit at rank 4.2 on average---near the edge of the typical top-$k$=5 window---and are rarely top-ranked, leaving the LLM to discriminate among lexically similar candidates; lexical canonicalization sharpens the gold-entry rank and largely eliminates this discrimination burden. For entity attribution, vanilla RAG sometimes \textit{degrades} performance below closed-book baselines (e.g., GPT-5 ATA: .83$\rightarrow$.74; Gemini-2.5-Pro ATA: .63$\rightarrow$.56). This counterintuitive degradation arises because entity attribution's one-hop structure means each incorrectly retrieved taxonomy entry directly becomes a wrong answer element, actively harming precision rather than merely failing to help. DtR addresses this by decomposing narratives into atomic behaviors before retrieval, improving precision from 34.2\% to 71.8\%.

\paragraph{Model Rankings}
Figure~\ref{fig:radar_overall}~(b) shows that GPT-5 leads at 81.4\%, while the next three models (Qwen-3-235B, GPT-4o, Claude-Sonnet-4) cluster within a narrow 1\% band (77.0--77.8\%). Importantly, no single model dominates all categories: Qwen-3-235B (open-source) tops entity linking with perfect F1 on three tasks, while GPT-5 leads entity attribution by 25 points over Qwen-3-235B (ATA DS: .90 vs.\ .65). This task-specific specialization confirms that CTIConnect discriminates along multiple capability axes rather than measuring a single dimension.


\subsection{Detailed Analysis}
\label{sec:findings}

\subsubsection{Semantic Gap Analysis}
\label{sec:findings-semantic-gap}

\begin{figure}[t]
    \centering
         \vspace{-0.2cm}
    \includegraphics[width=\linewidth]{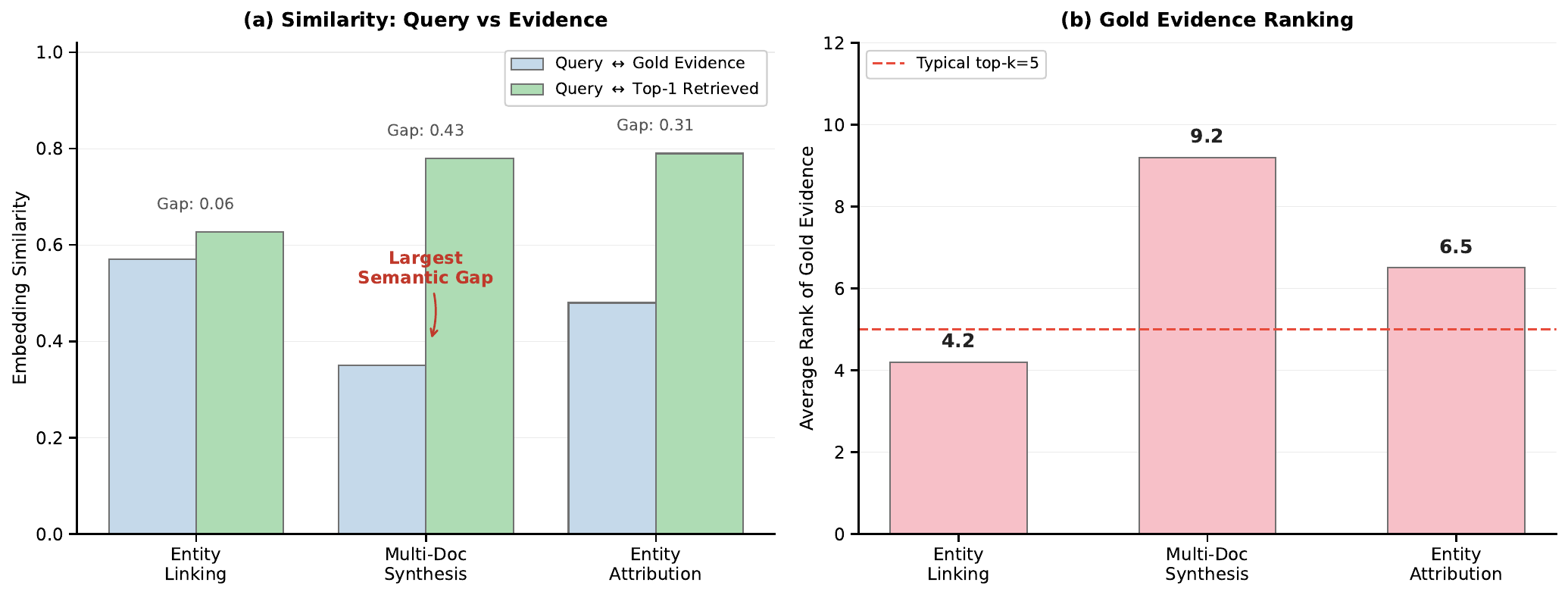}
     \vspace{-0.5cm}
    \caption{Semantic gap analysis. \textbf{(a)}~Query-to-gold vs.\ query-to-top-1 cosine similarity by task category. \textbf{(b)}~Average rank of gold evidence in vanilla RAG results; dashed line indicates the typical top-$k$=5 retrieval window.}
    \label{fig:semantic_gap}
    \vspace{-0.4cm}
\end{figure}

To understand \textit{why} vanilla RAG fails on certain task categories, we quantify the cross-source semantic gap, defined as the difference between query-to-gold and query-to-top-1 cosine similarity. As shown in Figure~\ref{fig:semantic_gap}, the left panel reveals \textbf{a clear gradient across task categories}: entity linking exhibits a small gap (0.06), indicating that gold evidence and top-retrieved entries are closely matched in embedding space. Entity attribution shows a substantially wider gap (0.31), reflecting the vocabulary mismatch between analyst-authored narratives and formal taxonomy terminology. Multi-document synthesis exhibits the widest gap (0.43), driven by two compounding effects on vendor reports. First, \textbf{entity aliasing}: the same threat actor, malware, or campaign appears under inconsistent names across vendors (e.g., ``APT29'' vs.\ ``Cozy Bear''), so embedding similarity often misses relevant reports that use a different alias than the query. Second, \textbf{chunk-level noise}: non-content elements (ads, navigation, boilerplate, generic threat-landscape filler) interleave with the actual intelligence and dilute chunk embeddings, letting lexically similar but topically unrelated distractors outscore gold documents.

The right panel reveals \textbf{how the gap manifests in retrieval ranks}. Entity linking gold entries rank 4.2 on average (near the edge of the typical top-$k$=5 window), so retrieval mostly succeeds; the small embedding gap instead surfaces as a candidate-discrimination difficulty, with the LLM forced to choose among lexically similar near-neighbors. Entity attribution gold entries fall to rank 6.5, just outside the retrieval window, because narrative descriptions and taxonomy entries use fundamentally different linguistic registers. Multi-document synthesis gold entries rank worst at 9.2: the aliasing and chunk-noise effects identified above jointly push gold beneath many near-miss distractors.
This two-dimensional analysis (gap severity \textit{and} retrieval rank) motivates the design of \textbf{distinct retrieval strategies for each task category}: canonicalization for entity linking, decomposition plus canonicalization for entity attribution, and entity resolution for multi-document synthesis.

\subsubsection{Evidence Utilization Analysis}

\begin{figure}[t]
    \centering
           \vspace{-0.2cm}
    \includegraphics[width=\linewidth]{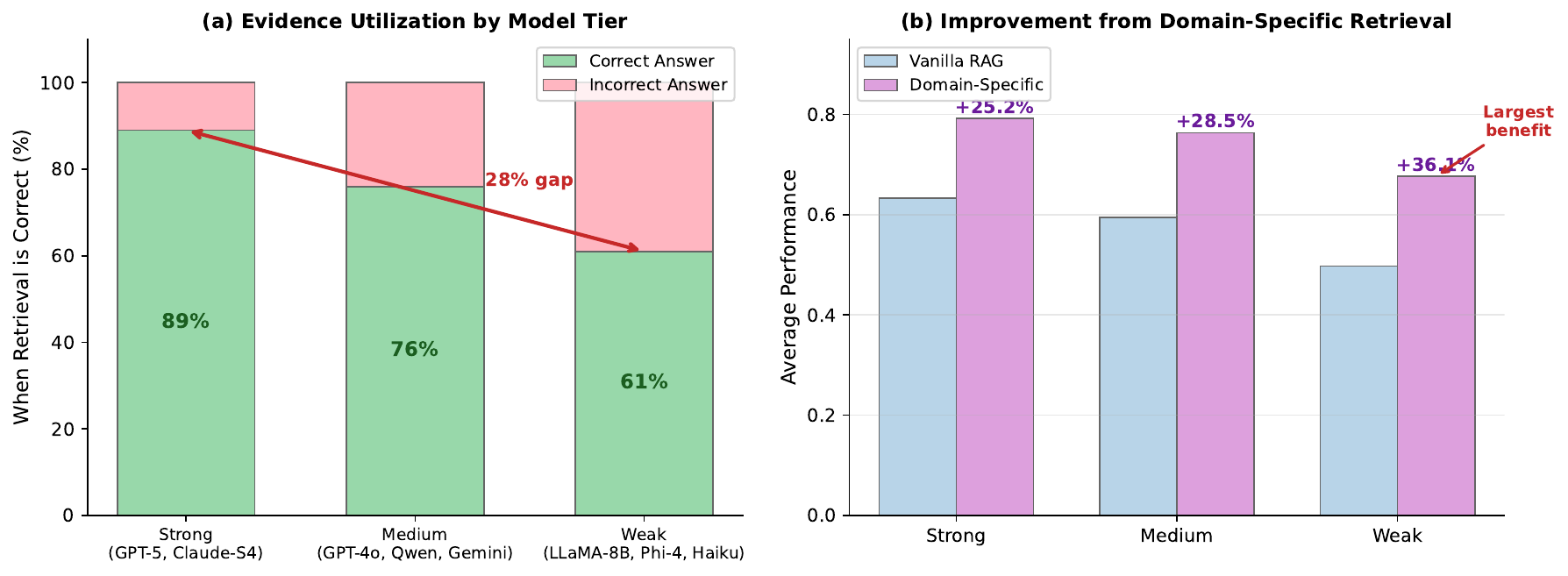}
       \vspace{-0.5cm}
\caption{Evidence utilization analysis across 100 selected instances. \textbf{(a)}~F1 conditioned on retrieval correctness, grouped by model tier. \textbf{(b)}~Performance improvement from domain-specific over vanilla RAG retrieval, by model tier.}
    \label{fig:model_utilization}
    \vspace{-0.3cm}
\end{figure}

Beyond retrieval strategy, we examine how models \textit{utilize} retrieved evidence by analyzing performance conditioned on retrieval quality across 100 selected instances. As shown in Figure~\ref{fig:model_utilization}~(a), when retrieval returns correct evidence, \textbf{a 28-point utilization gap} separates strong-tier models (GPT-5, Claude-Sonnet-4: 89\%) from weak-tier models (LLaMA-3-8B, Phi-4: 61\%), indicating that evidence utilization, not just evidence retrieval, is a bottleneck for smaller models. When retrieval returns incorrect evidence, strong models partially recover via parametric knowledge (GPT-5: 42\%) while weak models collapse (LLaMA-3-8B: 18\%); this asymmetry is especially damaging for entity attribution, where each incorrect retrieval directly introduces a wrong answer element.

Figure~\ref{fig:model_utilization}~(b) reveals a complementary pattern: domain-specific retrieval disproportionately benefits weaker models (weak-tier: +36.1\%; strong-tier: +25.2\%), such that \textbf{weak models with domain-specific retrieval nearly match strong models with vanilla RAG}. This finding suggests a practical trade-off for system deployment: for resource-constrained settings, investing in domain-specific retrieval infrastructure yields larger returns than scaling up model size with generic retrieval.

\begin{table}[t!]
\centering
\small
\caption{Intra-family scaling ($\Delta$\%) by task category. Scaling benefits concentrate in entity attribution, where each retrieval hop directly determines answer quality.}
\label{tab:scaling}
\vspace{-0.2cm}
\begin{tabular}{l|cc|ccc}
\toprule
\textbf{Family} & \textbf{Small} & \textbf{Large}
& \textbf{EL} & \textbf{MDS} & \textbf{EA} \\
\midrule
LLaMA-3    & 62.7 & 75.6  & +8.0  & +8.3  & +29.5 \\
Claude     & 69.9 & 77.0  & +4.2  & +1.7  & +21.0 \\
Gemini-2.5 & 74.1 & 76.4  & +0.3  & +0.7  & +9.0  \\
GPT        & 77.1 & 81.4  & +1.0  & +0.0  & +17.5 \\
\midrule
\textbf{Avg} & 70.9 & 77.6 & +3.4 & +2.7 & +19.2 \\
\bottomrule
\end{tabular}
\vspace{-0.4cm}
\end{table}

\subsubsection{Model Scaling Analysis}

The preceding analyses focus on retrieval; we now examine whether model choice matters equally across task categories. As shown in Table~\ref{tab:scaling}, scaling from a smaller to a larger model within the same family yields \textbf{highly uneven gains}: entity attribution benefits six times more than entity linking or multi-document synthesis (+19.2\% vs.\ +3.4\% and +2.7\%).

This asymmetry follows directly from where the performance bottleneck lies in each category. For entity linking and multi-document synthesis, the bottleneck is retrieval infrastructure (whether the system retrieves the right evidence) rather than model reasoning. Once evidence is retrieved, even smaller models can perform the required canonicalization or synthesis adequately, so scaling model size adds little. For entity attribution, the bottleneck shifts to model capability: DtR requires the model itself to decompose narratives into atomic behaviors and canonicalize each into taxonomy-aligned vocabulary before retrieval. Every improvement in decomposition quality directly translates into better retrieval targets and thus higher F1, explaining why scaling benefits are largest here (LLaMA-3: +29.5\% from 8B to 405B).

This bottleneck analysis also explains the overall leaderboard structure: GPT-5's lead over Qwen-3-235B (81.4\% vs.\ 77.8\%) concentrates almost entirely in entity attribution (ATA DS: .90 vs.\ .65), while Qwen-3-235B matches or exceeds all proprietary models on entity linking and multi-document synthesis. The practical implication is that \textbf{model selection should be task-aware}: smaller open-source models suffice for retrieval-bottlenecked categories, while entity attribution demands stronger model capabilities.

\subsubsection{Comparison with General-Purpose Retrieval Improvements}
\label{sec:general_purpose_retrieval}

\begin{table}[t]
\small
\centering
    \vspace{-0.2cm}
\caption{Retrieval-paradigm comparison with GPT-4o as the default answering model. General-purpose improvements close only a small fraction of the VR$\rightarrow$DS gap.}
\vspace{-0.2cm}
\label{tab:rerank_ircot}
\begin{tabular}{l|l|c|cccc}
\toprule
\textbf{Category} & \textbf{Metric} & \textbf{CB} & \textbf{VR} & \textbf{Rerank} & \textbf{IRCoT} & \textbf{DS} \\
\midrule
\multirow{2}{*}{Entity Linking}
& Hit Rate & --   & .845 & .862 & .891 & .985 \\
& F1       & .03  & .677 & .708 & .732 & \textbf{.977} \\
\midrule
\multirow{2}{*}{Multi-Doc Synth.}
& Hit Rate & --   & .733 & .741 & .756 & .763 \\
& Judge    & --   & .533 & .542 & .553 & \textbf{.573} \\
\midrule
\multirow{2}{*}{Entity Attribution}
& Hit Rate & --   & .790 & .812 & .843 & .890 \\
& F1       & .53  & .610 & .623 & .635 & \textbf{.655} \\
\bottomrule
\end{tabular}
    \vspace{-0.4cm}
\end{table}

A natural question is whether the vanilla-to-domain-specific gap can be closed by recent general-purpose retrieval improvements. We evaluate two widely-adopted paradigms using \textbf{GPT-4o as the default answering model}: \emph{retrieve-then-rerank} (dense retrieval with \texttt{text-embedding-3-large} followed by reranking the top 50 candidates with the BAAI/bge-reranker-v2-m3 cross-encoder) and \emph{iterative retrieval} (IRCoT~\cite{ircot}; up to 3 rounds of CoT-guided query refinement). Both share the same encoder as vanilla RAG, isolating the effect of retrieval paradigm.
Table~\ref{tab:rerank_ircot} shows that Rerank and IRCoT improve over vanilla RAG by 1--5\% on average, consistently smaller than the VR$\rightarrow$DS gap of 5--30\%. This pattern confirms that the cross-source CTI semantic gap is \textbf{not closable by generic retrieval improvements}: it requires \textit{structural} interventions operating on the underlying vocabulary and entity-resolution mismatch (canonicalization, behavior decomposition, entity-graph routing) rather than incremental improvements to candidate ordering.

\subsubsection{Full-Benchmark Verification}
\label{sec:scalability}

Our primary multi-model evaluation (Table~\ref{tab:main_result}) is conducted on a 691-pair subset of \cticonnect{}, since running all ten models under three retrieval configurations---with GPT-4-judge scoring for the synthesis tasks---over the full benchmark is computationally costly. To confirm that these conclusions hold at full scale, we re-evaluate three representative models spanning performance tiers---one frontier proprietary (GPT-5), one mid-tier proprietary (Gemini-2.5-Flash), and one weak-tier (Claude-3.5-Haiku)---on the complete 1,860-pair benchmark under all retrieval configurations.
Three findings emerge from Table~\ref{tab:scalability}: (1)~the model ranking (GPT-5~$>$~Gemini-2.5-Flash~$>$~Claude-3.5-Haiku) is preserved; (2)~the relative difficulty ordering across categories (EL~$\gg$~MDS~$\approx$~EA) is preserved; (3)~per-category performance patterns are consistent. Absolute scores shift by only 1--3.5\% on the full set, reflecting its broader distribution of harder cases (e.g., rarer cross-reference mappings and TTP-evolution questions on contradictory reports) rather than any change in measured capability.

\subsubsection{Temporal Robustness}
\label{sec:temporal_robustness}

A core property of CTI data is its rapid temporal evolution: new vulnerabilities, threat actors, and adversary techniques emerge continuously. To validate that \cticonnect{} generalizes to continually evolving CTI rather than overfitting to a specific temporal window, we conduct per-task temporal-split analysis spanning 2008--2025 (full results in~\cref{app:temporal_split}). Multi-document synthesis tasks are inherently cross-temporal: CSC, TAP, and MLA exhibit average temporal spans of 1.8--3.2 years per QA, with TAP being 100\% cross-temporal. For the remaining tasks, we split at each task's median temporal cutoff and find that closed-book, vanilla RAG, and domain-specific performance remain stable across temporal halves (typical absolute differences fall within $\pm$2\%), indicating that retrieval-based conclusions transfer from the historical to the recent portion of the benchmark and, by extension, to continually arriving CTI data.

\begin{table}[t]
\small
\centering
\vspace{-0.2cm}
\caption{Full-benchmark verification on three representative models. Model ranking and per-category structure are preserved across the entire 1,860-pair benchmark.}
    \vspace{-0.2cm}
\label{tab:scalability}
\begin{tabular}{l|cc|c}
\toprule
\textbf{Model} & \textbf{Subset (691)} & \textbf{Full (1,860)} & \textbf{$\Delta$} \\
\midrule
GPT-5             & 81.4\% & 77.9\% & $-3.5$ \\
Gemini-2.5-Flash  & 74.1\% & 72.6\% & $-1.5$ \\
Claude-3.5-Haiku  & 69.9\% & 68.9\% & $-1.0$ \\
\bottomrule
\end{tabular}
\vspace{-0.3cm}
\end{table}

\section{Conclusion}
\label{conclusion}

We introduced \cticonnect{}, the first benchmark for evaluating retrieval-augmented LLMs across heterogeneous, multi-source CTI. The benchmark contains 1,860 high-quality QA pairs grounded in authoritative CTI sources, produced through a unified construction pipeline that preserves discriminative structure across scales. Experiments reveal that the cross-source semantic gap is not a uniform obstacle but manifests differently across task categories (minimal for entity linking with proper retrieval yet severe for attribution and synthesis), demanding fundamentally different retrieval strategies rather than a one-size-fits-all approach. Equally important, the performance bottleneck shifts between retrieval infrastructure and model reasoning depending on task category: smaller open-source models match frontier proprietary models on retrieval-bottlenecked tasks, while entity attribution remains gated by model capability. 
We further verified that general-purpose retrieval improvements (retrieve-then-rerank, IRCoT) close only a small fraction of the vanilla-to-domain-specific gap. These conclusions hold on the full benchmark and under temporal splits across 2008--2025.
These results point toward LLM-powered security intelligence platforms that dynamically route analyst queries to task-appropriate retrieval strategies across heterogeneous knowledge sources. More broadly, they pave the way for future agentic harness design in cyber threat intelligence, where the central challenge is equipping agents to effectively exploit the massive vertical knowledge of the domain.

\begin{acks}
This work is supported by the National Science Foundation under grant 2442171 and the Google Academic Research Award (GARA). Any opinions, findings, and conclusions made in this paper are those of the authors and do not necessarily reflect the views of the funding agencies.
\end{acks}

\bibliographystyle{ACM-Reference-Format}
\bibliography{reference}

\appendix
\clearpage
\newpage
\onecolumn

\section*{Appendix}

\section{Validation and Robustness Analyses}
\label{app:validation_robustness}

This section provides four validation and robustness analyses: LLM-as-a-Judge reliability, temporal robustness, CSKG robustness, and the Magniber case study. Our code and data are available at our project page: \href{https://cticonnect.github.io/}{\textcolor{purple}{cticonnect.github.io}}.

\subsection{LLM-as-a-Judge Reliability}
\label{app:judge_reliability}

Multi-document synthesis tasks employ GPT-4 as an automatic judge, scoring atomic claims extracted from model predictions against reference answers. To rigorously assess the judge's suitability, consistency, and freedom from systematic bias, we conduct three complementary experiments on a 20\% random sample of MDS instances comprising 115 atomic claims.

\paragraph{Setup}
Three CTI experts (each with $\geq$3 years of practitioner experience) independently scored the 115 claims, achieving an inter-human Cohen's $\kappa = 0.93$, indicating that the human reference is itself highly reliable.

\paragraph{Experiment 1: Human--LLM Agreement}
We computed Cohen's $\kappa$ between the GPT-4 judge and the expert-consensus labels. The overall agreement was $\kappa = 0.85$ (``almost perfect''), with per-task values of $\kappa_{\text{TAP}}=0.90$, $\kappa_{\text{MLA}}=0.85$, and $\kappa_{\text{CSC}}=0.81$.

\paragraph{Experiment 2: Self-Consistency}
We ran the GPT-4 judge five independent times on the same 115 claims. The judge produced \emph{identical} verdicts on 108/115 (93.9\%) claims across all five runs, with per-QA score standard deviation of 0.02. The judge is therefore stable rather than stochastic.

\paragraph{Experiment 3: Stylistic-Bias Detection}
A primary concern with LLM-as-a-Judge evaluation is that the judge may favor outputs sharing its own family's stylistic patterns. We test this by computing per-claim bias $= (\text{judge score} - \text{human score})$ across six representative models spanning five providers (Table~\ref{tab:judge_bias}).

\begin{table}[h]
\small
\centering
\caption{Per-model stylistic bias of the GPT-4 judge. GPT-family mean (+0.021) is comparable to non-GPT mean (+0.008); the differential $\Delta=0.013$ is negligible relative to score variance.}
\label{tab:judge_bias}
\begin{tabular}{l|l|r}
\toprule
\textbf{Model} & \textbf{Family} & \textbf{Mean Bias} \\
\midrule
GPT-5            & GPT     & +0.018 \\
GPT-4o           & GPT     & +0.024 \\
Gemini-2.5-Pro   & non-GPT & +0.031 \\
Claude-Sonnet-4  & non-GPT & $-0.008$ \\
LLaMA-3-405B     & non-GPT & +0.029 \\
Phi-4            & non-GPT & $-0.022$ \\
\midrule
GPT-family mean      & ---     & +0.021 \\
non-GPT mean         & ---     & +0.008 \\
$|\Delta|$           & ---     & 0.013  \\
\bottomrule
\end{tabular}
\end{table}

Notably, the highest per-model bias is observed for two non-GPT models, Gemini-2.5-Pro (+0.031) and LLaMA-3-405B (+0.029), both exceeding the GPT-4o bias (+0.024). This rules out a systematic preference of the GPT-4 judge for GPT-family outputs. The aggregate differential $|\Delta| = 0.013$ is approximately one-third of typical per-task standard deviation, and is therefore negligible for ranking decisions.

\paragraph{Summary}
Across the three experiments, the GPT-4 judge demonstrates (i)~strong agreement with expert humans, (ii)~stable self-consistency, and (iii)~no measurable systematic bias toward its own family. These findings support the validity of LLM-as-a-Judge scoring for the synthesis tasks in \cticonnect{}.

\subsection{Temporal Generalization}
\label{app:temporal_split}

\cticonnect{} spans CTI data from 2008 to 2025, enabling analysis of whether benchmark conclusions generalize to continually evolving CTI. Multi-document synthesis tasks are inherently cross-temporal by construction: 67\% of CSC QAs span multiple years (average span 1.8 yr), 100\% of TAP QAs are cross-temporal (3.2 yr), and 83\% of MLA QAs are cross-temporal (3.2 yr).

For entity linking and entity attribution tasks, we split each task at its per-task median temporal cutoff and re-evaluate under all three retrieval configurations (Table~\ref{tab:temporal_split}).

\begin{table}[h]
\small
\centering
\caption{Per-task temporal-split performance (older / newer halves). Differences across temporal halves are within $\pm$2\% across all settings, indicating stable generalization.}
\label{tab:temporal_split}
\begin{tabular}{l|c|c|c|c}
\toprule
\textbf{Task} & \textbf{Split (N)} & \textbf{CB} & \textbf{VR} & \textbf{DS} \\
\midrule
RCM & 31 / 69 & .016 / .000 & .968 / 1.00 & 1.00 / 1.00 \\
WIM & 49 / 51 & .000 / .000 & 1.00 / 1.00 & 1.00 / 1.00 \\
ATD & 32 / 68 & .04 / .04   & .72 / .73   & .99 / .99   \\
ESD & 50 / 50 & .00 / .00   & .66 / .68   & .98 / .98   \\
ATA & 24 / 36 & .63 / .58   & .67 / .62   & .72 / .67   \\
VCA & 26 / 64 & .46 / .49   & .55 / .59   & .59 / .63   \\
\bottomrule
\end{tabular}
\end{table}

Performance is stable across temporal halves under all three retrieval configurations, with typical absolute differences falling within $\pm$2\%. This stability indicates that the benchmark's discriminative properties transfer from the historical to the recent portion and, by extension, are expected to hold as CTI data continues to evolve.

\subsection{CSKG Robustness Analysis}
\label{app:cskg_robustness}

CSKG-Guided RAG (used for multi-document synthesis) constructs a Cybersecurity Knowledge Graph offline by extracting named entities from each corpus report. To clarify the dependence on the extraction model and provide a transparent cost analysis, this section documents (1) the model-dependency of CSKG quality and (2) the amortized construction cost.

\paragraph{Scope of Dependency}
Of the three domain-specific strategies, two (EtR and DtR) execute their query transformation using the \emph{model being evaluated itself}. A weaker answering model therefore naturally produces weaker transformations and lower end-to-end scores, with no external dependency. CSKG-Guided RAG is the only strategy whose offline graph is built once and shared across all evaluated models.

\paragraph{Robustness Experiment}
To isolate the impact of the CSKG-extraction model, we rebuilt the CSKG using GPT-4o-mini (in place of GPT-4o) and re-evaluated all three MDS tasks with GPT-4o as the answering model (Table~\ref{tab:cskg_robustness}).

\begin{table}[h]
\small
\centering
\caption{CSKG-Guided RAG with GPT-4o vs.\ GPT-4o-mini as the offline entity extractor. Average degradation is 1.6\%.}
\label{tab:cskg_robustness}
\begin{tabular}{l|cc|c}
\toprule
\textbf{Task} & \textbf{CSKG (GPT-4o)} & \textbf{CSKG (GPT-4o-mini)} & \textbf{$\Delta$} \\
\midrule
CSC & .660 & .645 & $-0.015$ \\
TAP & .670 & .645 & $-0.025$ \\
MLA & .390 & .383 & $-0.007$ \\
\midrule
\textbf{Avg.\ degradation} & --- & --- & $-0.016$ \\
\bottomrule
\end{tabular}
\end{table}

Average degradation is only 1.6\%, indicating that CSKG-Guided retrieval is robust to the underlying extraction model's capability. The robustness arises because retrieval scores reports by BM25 over shared canonical entities with IDF weighting, which depends primarily on whether each canonical entity is extracted at least once, not on subtle differences in extraction phrasing.

\paragraph{Cost Analysis}
CSKG construction is a one-time offline cost amortized across all subsequent queries. Per report, CSKG construction requires approximately 18K input tokens and 3K output tokens. For the full 321-report corpus, the total construction cost is $\sim$\$24 with GPT-4o and $\sim$\$1.4 with GPT-4o-mini. By comparison, the dominant cost in deployment remains answer-generation, performed once per query by the evaluated model itself. CSKG construction is therefore a negligible fraction of total operating cost.

\subsection{Magniber Cross-Source Correlation Case Study}
\label{app:magniber_case_study}

We illustrate the operational value of cross-source CTI correlation via the Magniber ransomware family, whose six-year operational continuity is invisible to single-report analysis.

\paragraph{Input}
A 2023-03 Google TAG report describing Magniber ransomware actors exploiting a variant of the Microsoft SmartScreen bypass.

\paragraph{CSKG-Guided Retrieval Output}
Running CSKG-Guided RAG on the 2023-03 input report surfaces two earlier reports that share many canonical entities with it:
\begin{itemize}[leftmargin=1.2em]
    \item ThreatPost (2017-10): documents an earlier Magniber variant targeting South Korean users.
    \item CrowdStrike (2021-08): describes Magniber's use of the PrintNightmare vulnerability (CVE-2021-34527) for victim infection.
\end{itemize}

\paragraph{Synthesized Intelligence}
Linking the three reports reveals a six-year (2017--2023) Magniber campaign with consistent malware family identity but rotating delivery vectors: geographic targeting evolution (South Korea $\rightarrow$ broader Windows users), shifting exploitation chains (initial dropper $\rightarrow$ PrintNightmare $\rightarrow$ SmartScreen bypass), and persistent operational infrastructure. None of the three reports alone establishes this continuity; vanilla RAG fails to surface the 2017 and 2021 reports because vendor-specific naming variations and changing exploit terminology displace the gold reports below ranking thresholds. This case demonstrates the value of cross-source correlation: it surfaces patterns that remain invisible to manual, single-report analysis.

\end{document}